\documentclass[12pt,preprint]{aastex}

\slugcomment{Astrophysical Journal Letter}

\shorttitle{UGC 5101 X-ray}
\shortauthors{Imanishi et al.}

\begin{document}

\title{X-ray Evidence for a Buried Active Galactic Nucleus in UGC 5101}

\author{Masatoshi Imanishi \altaffilmark{1}, Yuichi Terashima
\altaffilmark{2}, Naohisa Anabuki \altaffilmark{2}, and Takao Nakagawa
\altaffilmark{2}}  

\altaffiltext{1}{National Astronomical Observatory, 2-21-1, Osawa,
Mitaka, Tokyo 181-8588, Japan; imanishi@optik.mtk.nao.ac.jp}
\altaffiltext{2}{The Institute of Space and Astronautical Science, 3-1-1
Yoshinodai, Sagamihara, Kanagawa 226-8510, Japan}

\begin{abstract}
We present X-ray observations of the ultraluminous infrared galaxy,
UGC 5101, thought to contain a buried active galactic 
nucleus (AGN) based on observations in other wavebands.
We detected an absorbed hard component at $>$3 keV, as well as soft
emission in the energy range 0.5--2 keV. 
The soft X-ray component, possibly due to a modestly dust-obscured,
extended starburst, has an absorption-corrected 0.5--2 keV X-ray
luminosity of L$_{X}$(0.5--2 keV) = 1.2 $\times$ 10$^{41}$ ergs s$^{-1}$. 
The 0.5--2 keV X-ray to infrared luminosity ratio is a factor of
$\sim$5 lower than typical values for a normal starburst,
suggesting that this extended starburst is unlikely to be energetically
dominant in UGC 5101.  
The most plausible origin of the absorbed hard component is the putative 
buried AGN. 
The 6.4 keV Fe K$\alpha$
emission line has a modest equivalent width ($\sim$400 eV), suggesting that
this hard component 
is direct emission from the AGN, rather than a scattered component. 
The absorption-corrected 2--10 keV X-ray luminosity of the buried AGN was 
estimated to be L$_{X}$(2--10 keV) $\sim$ 5 $\times$ 10$^{42}$ ergs s$^{-1}$.
The intrinsic 2--10 keV X-ray luminosity and the 2--10 keV
X-ray to infrared luminosity ratio are both comparable to values
measured for Mrk 463, a Seyfert-2 galaxy of similar infrared luminosity.  

\end{abstract}

\keywords{galaxies: nuclei --- X-ray: galaxies -- galaxies: individual
(UGC 5101)} 

\section{Introduction}

Ultraluminous infrared galaxies (ULIRGs) detected by the {\it IRAS}
all-sky survey radiate large, quasar-like luminosities 
($>$ a few $\times$ 10$^{45}$ ergs s$^{-1}$)
as infrared dust emission \citep{san88}; therefore they must
consist of extremely powerful energy sources that are obscured by dust.  
ULIRGs have also been found at large distances ($z >$ 0.3), where they have 
been used to trace the dust-obscured star formation rate, dust content,
and metallicity of the early universe \citep{bla99}. 
However, it remains unclear whether ULIRGs are powered by dust-obscured
active galactic nuclei (AGNs) or starbursts. 

It is now widely accepted that the energetically dominant components in
the majority of ULIRGs are compact ($<$a few 100 pc), highly 
dust-obscured nuclear energy sources, rather than extended
($\sim$kpc), weakly obscured starburst activity
\citep{soi00}.
It is essential to determine whether these compact energy sources consist of
AGNs and/or compact starbursts. 
Approximately 30\% of ULIRGs classified optically as Seyferts are known
to possess energetically important AGNs surrounded by dust with the
geometry of a {\it torus}  
\citep{vei99a,vei99b}.
However, the AGNs expected to reside in the majority of non-Seyfert 
ULIRGs may be deeply buried in dust along all lines of sight. 
It is necessary to determine the energetic importance of such 
{\it buried} AGNs  in order to understand the true nature of ULIRGs. 

To estimate the relative energetic contributions of an AGN and a
starburst, it is necessary to utilize signatures that are produced by only
one of these processes. 
One such signature is the polycyclic aromatic hydrocarbon (PAH)
emission found in the infrared 3--15 $\mu$m wavelength range
\citep{all89}. 
PAH emission is usually observed in a normal starburst, but not
detected in a pure AGN \citep{moo86,roc91}. 
Thus, PAH emission can be used to detect starburst emission, by
disentangling from AGN emission, in an AGN/starburst composite, such as
a ULIRG, and PAH luminosity traces the starbutst's absolute magnitude
\citep{mou90,gen00,ima02}.  
The 3.3 $\mu$m PAH emission feature and dust absorption
features found in infrared 3--4 $\mu$m spectra have been used to detect
candidates for non-Seyfert ULIRGs possessing energetically important
buried AGNs \citep{imd00,ima01,ima03}. 

One example of this type of object is UGC 5101 (z = 0.040). 
It has an infrared (8--1000 $\mu$m) luminosity of L$_{\rm IR}$ $\sim$3
$\times$ 10$^{45}$ ergs s$^{-1}$, a cool far-infrared color ({\it IRAS} 
25$\mu$m-to-60$\mu$m flux ratio of $<$0.2), and is optically classified
as a LINER, based on a systematic spectral classification of
galaxies \citep{vei95}.    
Imanishi et al. (2001) found the observed infrared 3--4 $\mu$m spectrum of
UGC 5101 to be best explained by the presence of a powerful buried AGN. 
\citet{ima03} argued that the nuclear energy source in UGC 5101 
is more centrally concentrated than nuclear gas and dust, as is expected
for a buried AGN, rather than a normal starburst, which shows a mixed
dust geometry \citep{soi00}.  
Radio VLBI observations \citep{smi98,lon03} and a near-infrared 
high spatial resolution image obtained with {\it HST} \citep{sco00} also
detected compact, high surface-brightness emission, which was
interpreted as an AGN by these authors. 
Therefore, there is mounting evidence for the presence of a buried AGN
in this galaxy. 
The detection of strong, absorbed 2--10 keV X-ray emission can provide 
compelling evidence for a powerful buried AGN, because an AGN is a much
stronger 2--10 keV X-ray emitter than a starburst.
Previous X-ray observations with {\it Chandra} failed to
find clear evidence for X-ray emission from the putative AGN
\citep{pta03}.  
Here, we report analysis of high quality X-ray
data of UGC 5101 obtained with {\it XMM} and {\it Chandra}.
The high spectral sensitivity of {\it XMM} at 2--10 keV was fully
utilized to detect absorbed hard X-ray emission from the putative
buried AGN. 
{\it Chandra} data are used to investigate the X-ray morphology,
benefiting from its high image quality, as well as to check the overall
spectral consistency with the {\it XMM} spectrum.  
Throughout this Letter, we use a cosmology of H$_{0}$ $=$ 75 km s$^{-1}$
Mpc$^{-1}$, $\Omega_{\rm M}$ = 0.3, and $\Omega_{\rm \Lambda}$ = 0.7.  

\section{Observations and Data Analysis}

An {\it XMM} observation of UGC 5101 was performed with the EPIC cameras
(PN, MOS1, and MOS2) in full-frame mode on 2001 November 12. 
We used standard data analysis procedures with the XMM Science
Analysis Software (SAS version 5.4), together with standard software packages
(FTOOLS 5.2 and XSPEC 11.2). 
Time intervals with high background count rates were removed from the
events files, and we included only events corresponding to patterns 0--4 for
PN and 0--12 for MOS, as suggested in the SAS Handbook.  
The resulting exposure times, after screening for time intervals with
high background, were 25.1 and 33.1 ksec for PN and MOS, respectively. 
We used the most recent calibration files retrieved from the XMM web page and 
response matrices created with SAS for our data analysis.  

We extracted a source spectrum using a 22.5 arcsec radius circular
region. 
A background spectrum was extracted from a nearby source-free circular
region with a radius of 60 arcsec. 
We binned the spectra so that each energy channel contained a minimum of 20
counts. 
An absorbed hard X-ray component was clearly detected only in the PN
spectrum, which contains the most counts. 
Therefore, we used the PN spectrum to obtain rough measurements of
important parameters.  
However, to improve the statistics, final spectral fitting was performed 
using the combined PN, MOS1 and MOS2 spectra. 
We allowed the relative normalizations of the cameras to vary in our
spectral fits as there is some uncertainty in the relative flux
calibration of the three cameras.

A {\it Chandra} observation was performed on 2001 May 28 with the
ACIS-S3 back-illuminated CCD chip.  
The data were reprocessed using CIAO 2.2.1 and CALDB 2.17. 
An effective exposure time of 47.8 ksec was obtained after discarding 
high background intervals. 

A {\it Chandra} spectrum of the nucleus was extracted from a circular
region with a radius of 8 arcsec. 
A background spectrum was extracted from an annular region around the
nucleus and subtracted from the nuclear spectrum.
The spectrum was binned to facilitate the detection of emission-line features.
The $C$-statistic was used to fit the spectrum since the number of
events in each bin is small.  

\section{Results}

\begin{figure}
\plotone{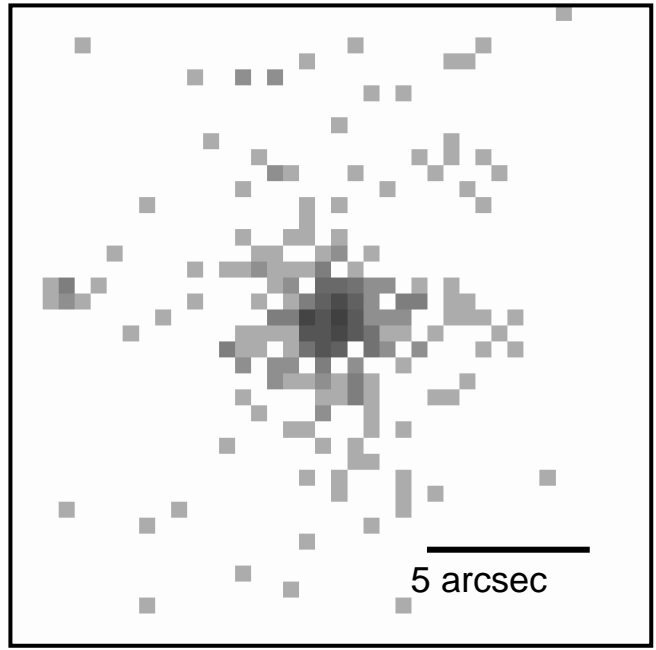}
\caption{A {\it Chandra} image of UGC 5101 in the 0.5--2 keV X-ray band.
\label{fig1}}
\end{figure}

Figure~\ref{fig1} shows a 0.5--2 keV {\it Chandra} image. 
The source is extended beyond the PSF of {\it Chandra} ($\sim$1 arcsec
diameter for $>$85\% encircled energy).  

\begin{figure}
\plottwo{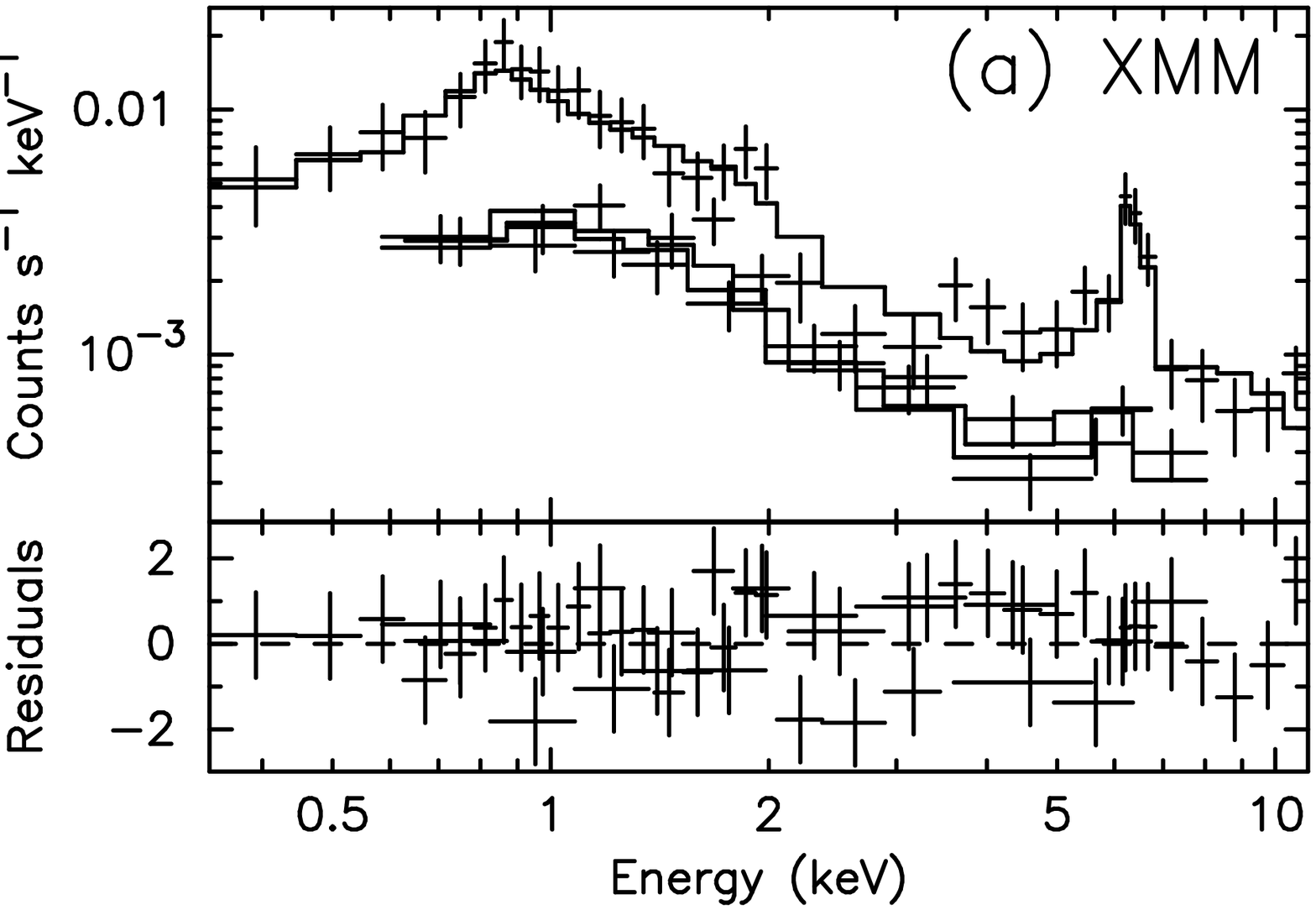}{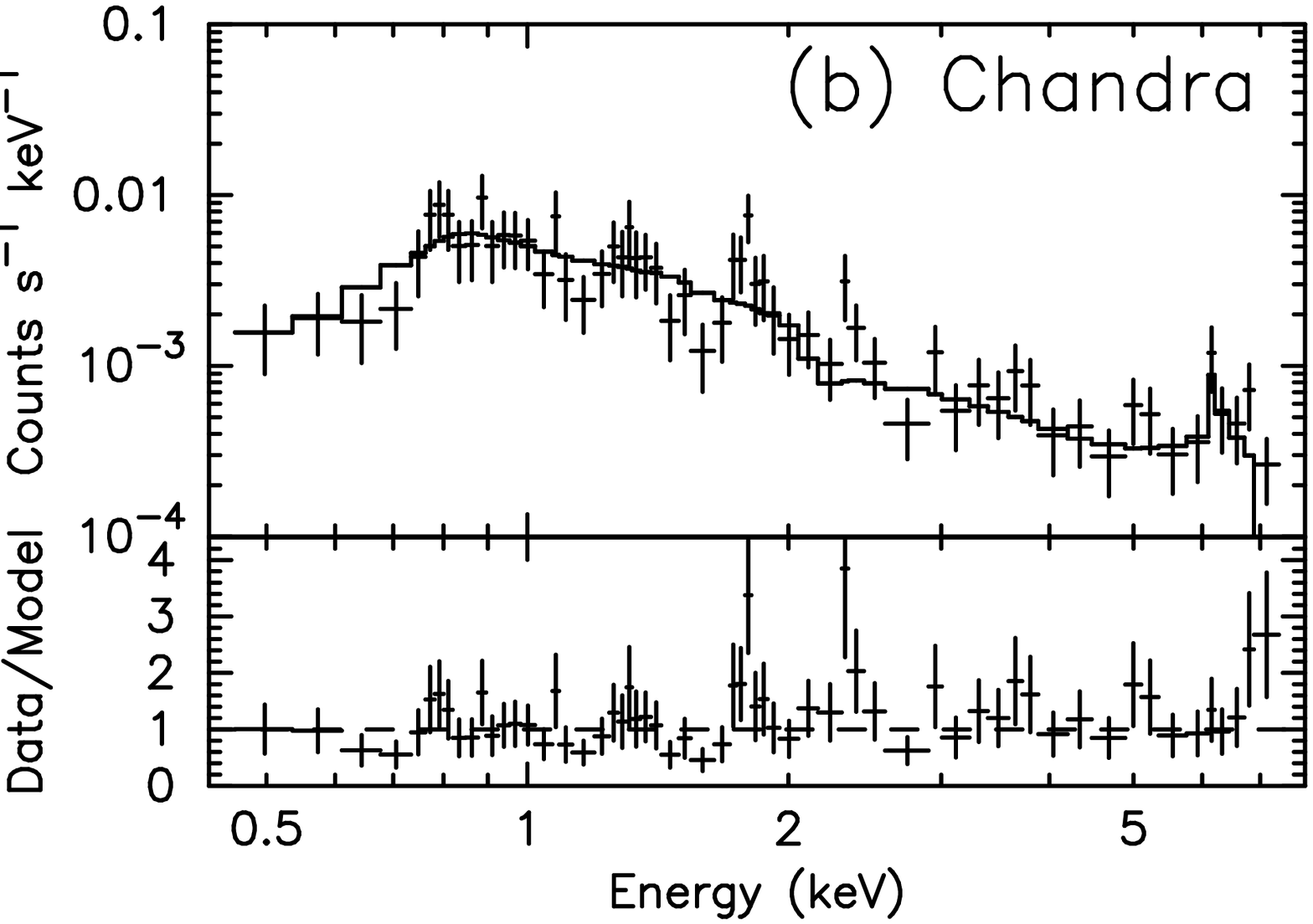}
\caption{
{\it (a)}: An {\it XMM} EPIC (PN, MOS1, and MOS2) spectrum of UGC 5101,
with model A in Table 1 overplotted. 
In the upper panel, the higher plots are a PN spectrum (which contains
the most counts), and the lower plots are MOS spectra.  
Residuals of the data from the best-fit model are displayed in the lower 
panel. 
{\it (b)}: A {\it Chandra} ACIS spectrum of the nuclear region.
The model A for {\it XMM} EPIC data is overplotted as a solid line.
The lower panel shows the ratio between the data and the model.
\label{fig2}}
\end{figure}

Figure~\ref{fig2}a shows the {\it XMM} EPIC spectrum of UGC 5101. 
In addition to the soft emission dominating the 0.5--2 keV 
X-ray range, an absorbed hard component at $>$3 keV and 
emission line at $\sim$6.5 keV were clearly seen. 
The hard component is likely to originate in the putative buried
AGN, and so we fitted this component with a power-law model. 
The emission line at $\sim$6.5 keV is likely to be the neutral Fe K$\alpha$
line at 6.4 keV, and was fitted with a narrow Gaussian profile. 
For the soft component, a 0.5--1 keV thermal emission model (mekal
model in XSPEC) from a starburst was first attempted to fit the 
soft X-ray spectrum, particularly the apparent spectral peak 
at 0.7--1 keV.  
However, significant residuals were present in this model fit. 
Therefore, we added 
(A) higher temperature thermal emission (5--10 keV) from hot gas
and X-ray binaries in a starburst, or (B) a second power-law component,
which could be due to Thomson scattering of the power-law emission from the
AGN.  
Table~\ref{tab1} summarizes the 
best-fit parameters for these models. 
To estimate the power of the putative
buried AGN, the most important measurement is the absorption-corrected
2--10 keV X-ray luminosity of the hard component, L$_{X}$(2--10 keV). 
Both models result in a similar L$_{X}$(2--10 keV) of $\sim$5 $\times$
10$^{42}$ ergs s$^{-1}$ and have a 6.4 keV Fe
K$\alpha$ line of modest equivalent width (EW$_{6.4}$ $\sim$ 400 eV).  
We varied some of the model parameters within reasonable ranges 
(e.g., a power-law photon index of $\Gamma$ = 1.0--2.0, a temperature
between 5--9 keV for the high-temperature thermal emission, and
abundances of 0.1--0.4 solar for the thermal emission); however, these
changes result in values of L$_{X}$(2-10 keV), which are consistent
within a factor of $<$2.  
 
Figure~\ref{fig2}b shows the {\it Chandra} ACIS spectrum. 
The solid line overplotted for comparison is based on model A (in Table 1)
from the {\it XMM} EPIC spectrum, with normalization of 
the continuum and of the Gaussian component representing the Fe K$\alpha$ line
as free parameters and were determined from the 
{\it Chandra} ACIS spectrum.   
The overall spectral shape is very similar to the {\it XMM} spectrum.
In the soft energy band, several emission lines are seen: strong He-like
Si ($\sim$1.9 keV) and  He-like S ($\sim$2.5 keV). 
We detected the absorbed hard component, but the amount of absorption of
this component is not well constrained from the {\it Chandra} spectrum
alone, because of the paucity of photons at the highest energies ($>$7
keV).  

The Fe K$\alpha$ emission feature at 6.4 keV is also detected in the 
{\it Chandra} spectrum. 
To estimate its equivalent width, we fitted the {\it Chandra} spectrum in the
4--7.5 keV band to minimize uncertainties in the modeling
of the continuum. 
Model A (in Table 1) was applied to the continuum. 
The equivalent width was estimated to be EW$_{6.4}$ =
$230^{+320}_{-230}$ eV ($\Delta \chi^2$ = 2.7).
This value is similar to the estimate based on the
{\it XMM} EPIC spectrum (Table 1). 
The extremely large equivalent width (5.9$^{+624}_{-3.2}$ keV, $\Delta
\chi^2$ = 4.605) reported by \citet{pta03} was not confirmed with our
analysis. 

\section{Discussion}

\subsection{Energetic Importance of Extended Starburst Activity}
 
UGC 5101 shows starburst activity with spatial extension of a few kpc
($\sim$4 arcsec) in the east-west direction \citep{gen98}, with dust
obscuration at a level similar to that of the nearby
prototypical starburst galaxy M82 in both mixed-dust and
foreground-screen dust models \citep{mcl93,stu00}.   
The extended morphology seen in the {\it Chandra} 0.5--2 keV image 
($\sim$4 arcsec in the east-west direction; Fig.1) 
and the detection of emission lines in our spectra (Fig.2) 
suggest that thermal emission from the extended starburst contributes
significantly to the soft X-ray emission below 2 keV.  
Assuming that all of the soft X-ray emission is of thermal 
origin (model A in Table 1), the absorption-corrected 0.5--2 keV 
soft X-ray luminosity is L$_{\rm X}$(0.5--2 keV)
$\sim$ 1.2 $\times$ 10$^{41}$ ergs s$^{-1}$, which results in  L$_{\rm
X}$(0.5--2 keV)/L$_{\rm IR}$ $\sim$ 4 $\times$ 10$^{-5}$. 

In a starburst galaxy, the 0.5--2 keV X-ray to far-infrared (40--500
$\mu$m) luminosity ratio is typically 
L$_{\rm X}$(0.5--2 keV)/L$_{\rm FIR}$ $\sim$ 2 $\times$ 
10$^{-4}$ \citep{ran03}.
As the far-infrared (40--500 $\mu$m) and infrared (8--1000 $\mu$m; 
Sanders \& Mirabel 1996) luminosities do not differ very much in an
ordinary starburst, the L$_{\rm X}$(0.5--2 keV)/L$_{\rm IR}$ ratio in
UGC 5101 is a factor of $\sim$5 lower than the value expected for a normal
starburst. 
If some fraction of the soft X-ray emission originates in 
scattering of the power-law AGN component, this discrepancy is  
even larger. 
In UGC 5101, the ratio of starburst-tracing 6.2 $\mu$m PAH luminosity 
(measured using large-aperture spectroscopy with 
{\it ISO}; Spoon et al. 2002) to infrared luminosity, is also lower
by a factor of $\sim$4 than that of a normal starburst \citep{ima03}. 
As the properties of the extended starburst in UGC 5101 are not expected
to differ significantly from a normal starburst, these results
both suggest that the modestly obscured extended starburst is unlikely to be
the dominant energy source in UGC 5101.

\subsection{A Buried Energy Source}

As the hard X-ray component is strongly absorbed,  
it cannot originate in X-ray binaries from the modestly
obscured extended starburst \citep{per02}.  
The most plausible origin is an AGN deeply
buried in gas and dust at the core.  
The modest equivalent width of the 6.4 keV Fe K$\alpha$ line suggests
that direct X-ray emission from 
the AGN is detected, because in physically plausible AGN geometries, 
a scattered X-ray component is expected to show a very large ($>$1 keV)
equivalent width \citep{mat03}.  
If the absorbed hard component is direct emission,
we find an absorption-corrected 2-10 keV luminosity for the buried AGN
of L$_{\rm X}$(2--10 keV) $\sim$ 5 $\times$ 10$^{42}$ ergs s$^{-1}$  
($\S$ 3).
In principle, the modest EW$_{6.4}$ of the detected hard component 
could be produced in a scattering situation, {\it if} it is Thomson
scattered by highly ionized gas. 
If this is the case, the intrinsic L$_{\rm X}$(2--10 keV) is even
higher than our estimate above. 

The L$_{\rm X}$(2--10 keV)/L$_{\rm IR}$ ratio for UGC 5101 is 
$\sim$2 $\times$ 10$^{-3}$.     
L$_{\rm X}$(2--10 keV) values have been measured for some Seyfert-2
 infrared luminous galaxies
(L$_{\rm IR}$ $>$ a few $\times $10$^{45}$ ergs s$^{-1}$) 
where the detected 2--10 keV emission is interpreted as a direct
component, because of the modest equivalent widths of the 6.4 keV Fe
K$\alpha$ emission ($<$1 keV).  
Examples include IRAS 20460+1925 \citep{oga97}, IRAS 23060+0505
\citep{bra97}, PKS 1345+12 \citep{ima99}, and Mrk 463 \citep{uen96}.
All are thought to be powered by AGNs, with no detectable starburst
activity in the infrared \citep{gen98,vei99b,imd00,ima02}. 
The L$_{\rm X}$(2--10 keV)/L$_{\rm IR}$ ratios of these known 
AGN-powered, Seyfert-2, infrared luminous galaxies are between 0.002 and
0.02, with Mrk 463  having the lowest value (L$_{\rm X}$(2--10 keV)
$\sim$ 4 $\times$ 10$^{42}$ ergs s$^{-1}$; Ueno et al. 1996, 
L$_{\rm IR}$ $\sim$ 2 $\times$ 10$^{45}$ ergs s$^{-1}$). 
A smaller value of L$_{\rm X}$(2--10 keV)/L$_{\rm IR}$ 
is expected for a buried AGN than for a Seyfert-2 AGN surrounded with
dust in a toroid geometry, because in the former case, a larger fraction
of the photons emitted from the AGN are absorbed by dust and re-emitted
in the infrared.   
Thus, the buried AGN in UGC 5101 is likely to be as powerful as that in
Mrk 463, in terms of both absolute 2--10 keV X-ray luminosity and  
ratio to infrared luminosity.  

Regardless of whether the primary energy source of UGC 5101 is a buried
AGN or a starburst, the large infrared luminosity is produced by UV emission.
The contribution from the detected X-ray-emitting buried AGN to the
total energy output in the infrared depends highly on the intrinsic UV
to 2--10 keV X-ray luminosity ratio; this is poorly understood and could
vary substantially among the population of AGNs.  
The buried AGN could account for the bulk of the infrared luminosity in
UGC 5101 if its intrinsic UV to 2--10 keV ratio is similar to the AGN
in Mrk 463.

By applying a template of infrared spectral energy distribution
of a {\it classical}, moderately infrared luminous AGN and a 
{\it normal} starburst to UGC 5101, \citet{far03} argued that infrared
dust emission flux from UGC 5101 at $\lambda$ $<$ 10 $\mu$m is dominated
by an AGN, while that at $\lambda$ $=$ 10--1000 $\mu$m is dominated by a
starburst.  
This conclusion comes primarily from the known fact that a classical AGN
generally shows a warmer infrared color than a normal starburst 
\citep{deg87}.  
However, ULIRGs are known to have a larger column density of
obscuring dust than classical, less infrared luminous AGNs
\citep{spo02,ima03}, and thus infrared emission from the buried AGNs in
ULIRGs should be cooler than that from classical AGNs.   
Furthermore, a significant fraction of dust grains in UGC
5101 was found to be ice-covered \citep{spo02,ima03}, which also causes
the infrared color to be cooler than bare dust grains \citep{dud03}.
Therefore, a much larger fraction of the infrared dust emission
from UGC 5101 at $>$10 $\mu$m can be produced by the buried AGN than that
estimated by Farrah et al..
Indeed, if an observed excess in the 60 $\mu$m luminosity function at
the highest luminosity end, over the extrapolation from lower
luminosity, is caused by AGN contribution \citep{tak03}, then it is
suggested that AGN-powered dust emission can contribute significantly to
the luminosities of ULIRGs at as long as 60 $\mu$m.   
These factors should be taken into account to correctly understand the
origin of strong dust emission from UGC 5101 at $\lambda$ $>$ 10 $\mu$m.

\acknowledgments

Y.T. and N.A. are supported by the Japan Society for the Promotion of
Science. The anonymous referee gave useful comments.

\clearpage

\begin{deluxetable}{ccccccc}
\tabletypesize{\small}
\tablecaption{Fitting results of the XMM data of UGC 5101. \label{tab1}}
\tablewidth{0pt}
\tablehead{
\colhead{Model} & \colhead{$\chi^{2}$/d.o.f.} & \colhead{kT} &
\colhead{N$_{\rm H}$} & \colhead{L$_{\rm TH}$(0.5--2keV)} & 
\colhead{L$_{\rm PL}$(2--10keV)} & \colhead{EW} \\
\colhead{} & \colhead{} & \colhead{[keV]} & \colhead{[10$^{23}$ cm$^{-2}$]} &
\colhead{[10$^{40}$ ergs s$^{-1}$]} & \colhead{[10$^{42}$ ergs
s$^{-1}$]} & \colhead{[eV]}  \\ 
\colhead{(1)} & \colhead{(2)} & \colhead{(3)} & \colhead{(4)} &
\colhead{(5)} & \colhead{(6)} & \colhead{(7)} 
}
\startdata
A & 45.7/46 & 0.7$^{+0.2}_{-0.3}$ & 14$^{+1}_{-2}$ & 12 & 7 &
410$^{+270}_{-240}$ \\ 
B & 50.5/46 & 0.6$^{+0.3}_{-0.4}$ & 12$^{+2}_{-1}$ & 3 & 6 &
330$^{+240}_{-200}$ \\  
\enddata

\vspace{1cm}

Note. --- 
Col. (1): Adopted best-fit models. 
Model A is absorbed power law
($\Gamma$ = 1.8 fixed) + 0.7 keV thermal + 10 keV (fixed) thermal +
narrow Gaussian (6.4 keV Fe K$\alpha$). 
Model B is absorbed power law ($\Gamma$ = 1.8 fixed) + 0.6 keV thermal + 
second power law ($\Gamma$ = 1.8 fixed; scattered component)+ narrow
Gaussian.  
In both models, the Galactic absorption is included. 
For thermal components, a mekal model, a fixed abundance of 0.5 solar, 
and plasma density of 1 cm$^{-3}$ were adopted. 
Col. (2): Reduced $\chi^{2}$ value.
Col. (3): Temperature for soft thermal emission.
The uncertainty is for 90\% confidence level for one parameter of
interest ($\Delta\chi^{2}$ = 2.7) throughout this Table.
Col. (4): Absorption for the main power law component. 
Col. (5): Absorption corrected 0.5--2 keV luminosity for thermal
emission. In model A, both 0.7 keV and 10 keV thermal components
were added. 
Col. (6): Absorption corrected 2--10 keV luminosity for the primary
power law component. 
Col. (7): Equivalent width of the 6.4 keV Fe K$\alpha$ emission line,
relative to the main power law component. 

\end{deluxetable}

\clearpage

\end{document}